\documentclass[10pt]{article}
\usepackage[dvips]{graphicx}
\usepackage{accents}
\begin{document}
\begin{center}
\textbf{{\large Statefinder Diagnostic for Born-Infeld Type Dark
Energy Model}}
 \vskip 0.35 in
\begin{minipage}{4.5 in}
\begin{center}
{\small Huang Zeng-Guang$^{1,\dag}$ and Lu Hui-Qing$^{2,\ddag}$
\vskip 0.06 in \textit{$^1$College of
Science,~Huaihai~Institute~of~Technology,~Lianyungang,~China
\\
$^\dag$zghuang@hhit.edu.cn
\\$^2$Department~of~Physics,~Shanghai~University,~Shanghai,~China
\\
$^\ddag$alberthq$\_$lu@staff.shu.edu.cn}}
\end{center}
\vskip 0.2 in

{\small Using a new method---statefinder diagnostic which can differ
one dark energy model from the others, we investigate in this letter
the dynamics of Born-Infeld(B-I) type dark energy model. The
evolutive trajectory of B-I type dark energy with Mexican hat
potential model with respect to $e-folding$ time $N$ is shown in the
$r(s)$ diagram. When the parameter of noncanonical kinetic energy
term $\eta\rightarrow0$ or kinetic energy
$\dot{\varphi}^2\rightarrow0$, B-I type dark energy(K-essence) model
reduces to Quintessence model or $\Lambda$CDM model corresponding to
the statefinder pair $\{r, s\}$=$\{1, 0\}$ respectively. As a
result, the the evolutive trajectory of our model in the $r(s)$
diagram in Mexican hat potential is quite different from those of
other dark energy models. \vskip 0.2 in \textit{Keywords:} Dark
energy; Born-Infeld; Statefinder; Attractor.
}
\end{minipage}
\end{center}
\vskip 0.2 in With the increase of the modern precise
cosmology[1-3], scientists believe more and more such a fact: the
Universe is nearly flat to high precision,
$\Omega_{total}=0.99\pm0.03$, and is undergoing a accelerated
expansion phase. To explain such a fact, an exoteric energy density
which is an unclumped form of energy density pervading the Universe
should be introduced. This unknown energy density called "dark
energy" with negative pressure, contributes to about two thirds of
the total energy density, and the density of clustered matter
including cold dark matters plus baryons contributes to about one
thirds of the total energy density. Perhaps the simplest explanation
for these data is that the dark energy corresponds to a positive
cosmological constant. However, there are two serious problems with
the cosmological constant, namely the "fine-tuning" and the cosmic
"coincidence". In the framework of quantum field theory, the
expectation value of vacuum energy is 123 order of magnitude larger
than the observed value of $10^{-47} GeV^4$. The absence of a
fundamental mechanism which sets the cosmological constant zero or
very small value is the first cosmological constant problem. The
second problem as the cosmic coincidence, states that why are the
energy densities of dark energy and dark matter nearly equal today?
An alternative is a scalar field which has not yet reached its
ground state.
\par It is well known that Hubble parameter $H(t)\equiv\frac{\dot{a}}{a}$ and deceleration
parameter $q_0$ are very important quantities which can describe the
geometrical properties of the Universe. However, both quantities
can't provide enough evidence to differentiate the more accurate
cosmological observational data and the more general models of dark
energy. For this reason, Sahni et al[4] propose a new geometrical
diagnostic pair $\{r, s\}$ for dark energy, which is called
statefinder and can be expressed as follows.
\begin{equation}r\equiv \frac{\dddot{a}}{aH^3},~~~~~~s\equiv\frac{r-1}{3(q-\frac{1}{2})}\end{equation}
Obviously, this diagnostic is constructed from the $a(t)$ and its
derivatives up to the third order. So, the statefinder probes the
expansion dynamics of the universe through higher derivatives of the
expansion factor. By far, many models[5] have been differentiated by
this geometrical diagnostic method. Its important property is that
$\{r, s\} = \{1, 0\}$ is a fixed point for the flat $\Lambda$CDM FRW
cosmological model. Departure of a given DE model from this fixed
point is a good way of establishing the "distance" of this model
from flat $\Lambda$CDM. In this letter, we will investigate the
evolutive trajectory of B-I type dark energy model in the $r-s$
diagram when the potential is taken as the Mexican hat potential,
and show the "distance" between our model and $\Lambda$CDM model.
\par Born and Infeld firstly introduce a nonlinear
electromagnetic field in 1934, and their original motivation is to
resolve the singularity in classical electromagnetic dynamics[6]. So
far, many authors have been studying the nonlinear B-I type string
theory and cosmology[7]. The lagrangian density for a B-I type
scalar field is
\begin{equation}\displaystyle L_S=\frac{1}{\eta}\left[1-\sqrt{1-\eta g^{\mu\nu}\varphi_{,~\mu}\varphi_{,~\nu}}~\right]\end{equation}
When parameter of noncanonical kinetic energy $\eta$ tends to be
zero, by Taylor expansion, Eq.(2) approximates to the lagrangian of
linear scalar field. This means that B-I type dark energy model will
reduce to Quintessence model if $\eta \rightarrow 0$.
\begin{equation}\lim_{\eta\rightarrow 0}L_S=\frac{1}{2}g^{\mu\nu}\varphi_{,~\mu}\varphi_{,~\nu}\end{equation}
Now we consider the Lagrangian with a potential $u(\varphi)$ in
spatially homogeneous scalar field, so, Eq.(2) becomes
\begin{equation}\end{equation}$$L_{ph}=\frac{1}{\eta}\left[1-\sqrt{1-\eta \dot{\varphi}^2}~\right]-u(\varphi)$$
From Eqs.(9) and (10), we know when kinetic energy
$\dot{\varphi}^2\rightarrow0$, B-I type dark energy model reduces to
$\Lambda$CDM model. In FRW space-time metric, Einstein equation
$G_{\mu\nu}=KT_{\mu\nu}$ can be written as
\begin{equation}H^2=\frac{1}{3}(\rho_\varphi+\rho_m)\end{equation}
\begin{equation}\dot{H}=-\frac{1}{2}(\rho_\varphi+p_\varphi+\rho_m)\end{equation}
\begin{equation}\dot{\rho}_m+3H\rho_m=0\end{equation}
\begin{equation}\dot{\rho}_\varphi+3H(\rho_\varphi+p_\varphi)=0\end{equation}
where $\rho_m$, $\rho_\varphi$ and $p_\varphi$ are the matter energy
densigy, the effective energy density and effective pressure of the
B-I type scalar field respectively, and we work in units $8\pi G=1$.
$\rho_\varphi$ and $p_\varphi$ can be expressed as follows
\begin{equation}\rho_\varphi=T^0_0=\frac{1}{\eta\sqrt{1-\eta \dot{\varphi}^2}}-\frac{1}{\eta}+u(\varphi)\end{equation}
\begin{equation}p_\varphi\equiv\omega_\varphi\rho_\varphi=-T^1_1=-T^2_2=-T^3_3=\frac{1}{\eta}-\frac{\sqrt{1-\eta \dot{\varphi}^2}}{\eta}-u(\varphi)\end{equation}
According to Eq.(1) and Eq.s(5-8), we can obtain the expression of
statefinder pair and deceleration parameter $q$
\begin{equation}r=1+\frac{9}{2}\Omega_\varphi\omega_\varphi(1+\omega_\varphi)-\frac{3}{2}\Omega_\varphi\omega'_\varphi\end{equation}
\begin{equation}s=1+\omega_\varphi-\frac{1}{3}\frac{\omega'_\varphi}{\omega_\varphi}\end{equation}
\begin{equation}q=\frac{1}{2}(1+3\Omega_\varphi\omega_\varphi)\end{equation}
where $\Omega_\varphi$ and $\omega_\varphi$ are parameter of energy
density and state equation respectively, and a prime sign denotes
the derivative with respect to the $"e-folding"$ time $N=lna$.
\par Mexican hat potential $u(\varphi)=\frac{\mu}{4}(\varphi^2-\varepsilon^2)^2+u_0$
with $\mu$, $\varepsilon$ and $u_0$ being constant, has been widely
studied in symmetry breaking problem of unification theory. In this
letter, we take this potential as the potential of B-I type scalar
field. \vskip 0.05 in
\begin{minipage}{0.5\textwidth}
\includegraphics[scale=0.8]{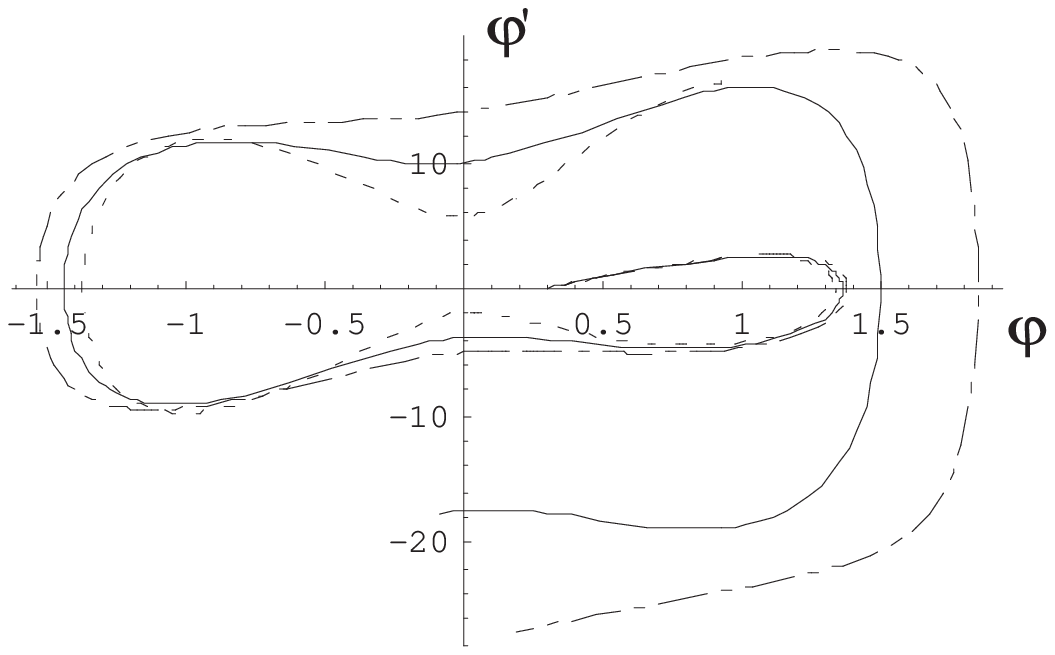}
{\small Fig.1 The attractive trajectories of B-I model with Mexican
hat potential are stable spirals for different $\eta=0.6$, 1, 1.3
respectively. We set $\mu=0.5$, $\varepsilon=0.95$, $u_0=1$.}
\end{minipage}
\hspace{0.02\textwidth}
\begin{minipage}{0.5\textwidth}
\includegraphics[scale=0.8]{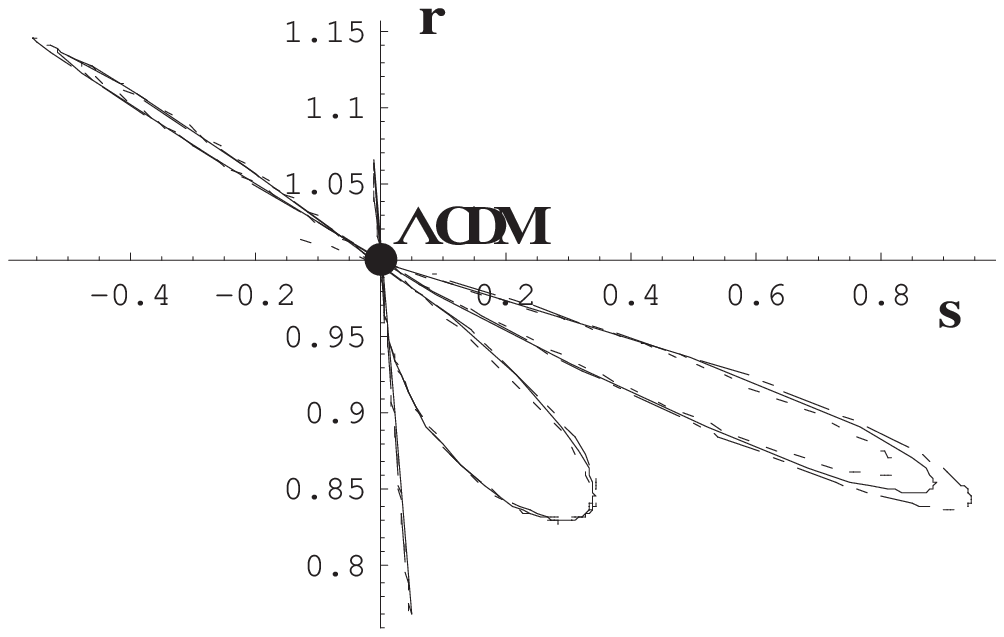}
{\small Fig.2 The evolutive trajectories of $r(s)$ for different
$\eta=1$(real line), 0.6(dot line), 1.3(dot-dashed line)
respectively.}
\end{minipage}

\begin{minipage}{0.5\textwidth}
\includegraphics[scale=0.8]{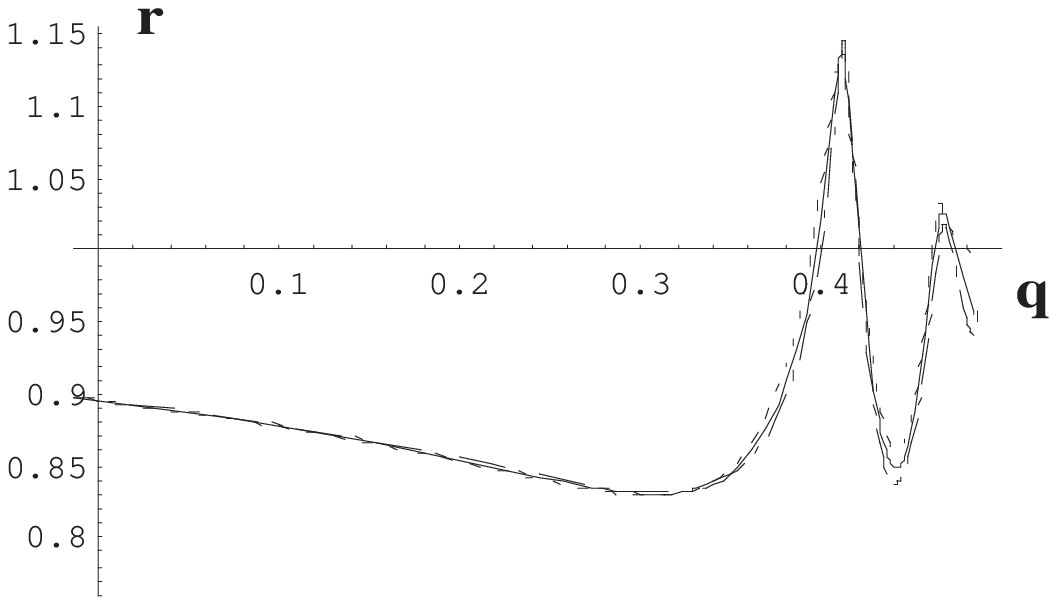}
{\small Fig.3 The evolutive trajectories of $r(q)$ for different
$\eta=1$(real line), 0.6(dot line), 1.3(dot-dashed line)
respectively.}
\end{minipage}
\hspace{0.02\textwidth}
\begin{minipage}{0.5\textwidth}
\includegraphics[scale=0.8]{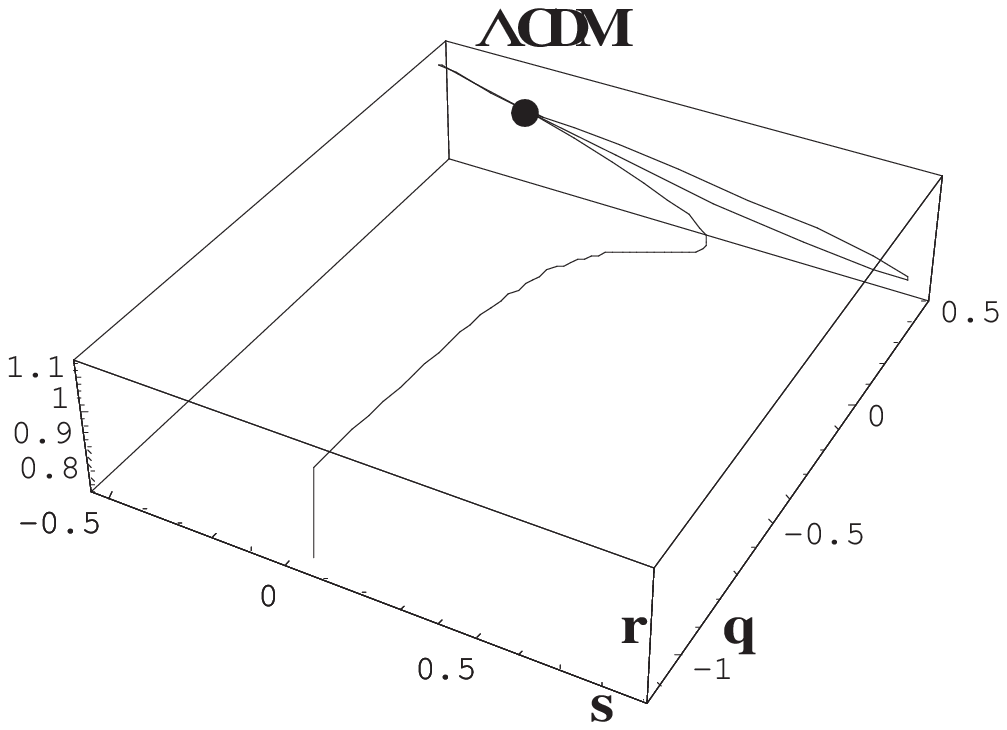}
{\small Fig.4 The 3D evolutive trajectories of $r-s-q$ for
$\eta=1$.}
\end{minipage}
\vskip 0.3 in
\par In summary, we investigate in this letter the dynamics of B-I type dark
energy model by using a new geometric diagnostic
method---statefinder pair $\{r,s\}$. We find that when
$\eta\rightarrow0$, B-I type dark energy(K-essence) model reduces to
Quintessence model with canonical kinetic energy term. The phase
portrait of $r-s$, $r-q$ and $r-s-q$ with respect to $e-folding$
time $N$ are shown mathematically. Fig.1 shows that the B-I type
dark energy model in Mexican hat potential admits a stable spiral
attractor solution corresponding to the dark energy density
$\Omega_\varphi\sim1$ and the equation of state parameter
$\omega_\varphi\sim-1$ meeting the current observations well. The
new geometric quantities $r$ and $s$ possesses some interesting
characters e.g. the evolutive trajectory of statefinder pair focuses
on a fixed point $\{r, s\}$=$\{1, 0\}$ corresponding to $\Lambda$CDM
model. Fig.2 shows that the evolutive trajectories of $r(s)$ will
always pass the fixed point $\{r, s\}$=$\{1, 0\}$ of $\Lambda$CDM
FRW cosmological model in the future when we set different value of
$\eta=0.6,1,1.3$ respectively. We can see easily the evolutive
trajectories form series swirls before reaches the attractor, and
are quite different from those of the other dark energy models whose
swirl is none or only one. The phase portrait of $r-q$ and 3D
parametric portrait of $r-s-q$ are shown in Fig.3 and Fig.4
respectively, where the parameter values are the same to those of
Fig.1. As a result, the evolutive behavior of B-I type dark energy
model in $r-s$ diagram will pass the fixed point corresponding to
$\Lambda$CDM model and is quite different from those of other dark
energy models.
\begin{flushleft}\textbf{Acknowledgements}\end{flushleft}
This work is partially supported by National Nature Science
Foundation of China under Grant No.10573012.

\begin{flushleft}{\noindent\bf References}
\small{
\item{1.}{ A. G. Riess, \textit{Astron. J}\textbf{116}, 1009(1998);
\\\hspace{0.15 in}S. Perlmutter et al., \textit{Astrophys. J}\textbf{517}, 565(1999).}
\item{2.}{ D. N. Spergel et al., Astrophys. J. Suppl\textbf{148}, 175(2003).}
\item{3.}{ P. de Bernardis et al., arXiv:astro-ph/0105296.}
\item{4.}{ V. Sahni, T. D. Saini, A. A. Starobinsky and U. Alam, \textit{JETP Lett}\textbf{77}, 201(2003).}
\item{5.}{ B. R. Chang, H. Y. Liu, L. X. Xu and C. W. Zhang, \textit{Chin. Phys. Lett.}\textbf{24}, 2153(2007);
\\\hspace{0.17 in}B. R. Chang, H. Y. Liu, L. X. Xu, C. W. Zhang and Y. L. Ping, \textit{JCAP}\textbf{0701}, 016(2007);
\\\hspace{0.17 in}Gorini, A. Kamenshchik and U. Moschella, \textit{Phys. Rev. D}\textbf{67}, 063509(2003);
\\\hspace{0.17 in}X. Zhang, \textit{Phys. Lett. B}\textbf{611}, 1(2005).}
\item{6.}{ M. Born and Z. Infeld, \textit{Proc. Roy. Soc. A}\textbf{144}, 425(1934).}
\item{7.}{ G. W. Gibbons and C. A. R. Herdeiro, \textit{Phys. Rev. D}\textbf{63}, 064006(2001);
\\\hspace{0.17 in}H. P. de Oliveira, \textit{J. Math. Phys}.\textbf{36}, 2988(1995);
\\\hspace{0.17 in}T. Taniuti, \textit{Prog. Theor. Phys.}(kyoto) Suppl\textbf{9}, 69(1958).}
}
\end{flushleft}
\end{document}